\documentclass[a4paper, 12pt]{extarticle}
\usepackage{graphicx}% Include figure files
\usepackage{adjustbox}
\usepackage{amsmath}
\usepackage{dcolumn}% Align table columns on decimal point
\usepackage{bm}% bold math
\usepackage{hyperref}
\usepackage{lineno}
\usepackage{xspace}                     
\usepackage{comment}
\usepackage{multirow}
\usepackage{cite}
\def\pythia\textsc{Pythia}
\newcommand{\pp}           {pp\xspace}

\newcommand{\ee}           {\mbox{$\mathrm {e^{+}e^{-}}$}\xspace}

\newcommand{\pt}           {\ensuremath{p_{\rm T}}\xspace}

\newcommand{\sqrts}        {\ensuremath{\sqrt{s}}\xspace}

\newcommand{\nineH}        {$\sqrt{s}=0.9$~Te\kern-.1emV\xspace}
\newcommand{\seven}        {$\sqrt{s}=7$~Te\kern-.1emV\xspace}
\newcommand{\twoH}         {$\sqrt{s}=0.2$~Te\kern-.1emV\xspace}
\newcommand{\twosevensix}  {$\sqrt{s}=2.76$~Te\kern-.1emV\xspace}
\newcommand{\five}         {$\sqrt{s}=5.02$~Te\kern-.1emV\xspace}
\newcommand{\sqrtsfive}{\five}
\newcommand{\thirteen}     {$\sqrt{s}=13$~Te\kern-.1emV\xspace}
\newcommand{\twosevensixnn}{$\sqrt{s_{\mathrm{NN}}}=2.76$~Te\kern-.1emV\xspace}
\newcommand{\fivenn}       {$\sqrt{s_{\mathrm{NN}}}=5.02$~Te\kern-.1emV\xspace}

\newcommand{\GeVc}{\ensuremath{\mathrm{GeV}/c}\xspace}
\newcommand{\MeVc}         {Me\kern-.1emV/$c$\xspace}
\newcommand{\TeV}          {Te\kern-.1emV\xspace}
\newcommand{\GeV}          {Ge\kern-.1emV\xspace}
\newcommand{\MeV}          {Me\kern-.1emV\xspace}
\newcommand{\GeVcc}        {Ge\kern-.2emV/$c^2$\xspace}
\newcommand{\MeVcc}        {Me\kern-.2emV/$c^2$\xspace}

\let\gevc=\GeVc

\let\tev=\TeV

\newcommand{\ccbar}{\ensuremath{c\overline{c}}\xspace}
\newcommand{\Dzero}        {\ensuremath{D^{0}}\xspace}
\newcommand{\Dplus}        {\ensuremath{D^{+}}\xspace}
\newcommand{\Dstar}        {\ensuremath{D^{*+}}\xspace}
\newcommand{\Ds}           {\ensuremath{D^{+}_{s}}\xspace}
\newcommand{\Lambdac}      {\ensuremath{\Lambda_{c}^{+}}\xspace}

\newcommand{\Xic}  {\ensuremath{ \Xi_{c}^{0,+}}\xspace}
\newcommand{\Xicp} {\ensuremath{ \Xi_{c}^{+}}\xspace}
\newcommand{\Xicz} {\ensuremath{ \Xi_{c}^{0}}\xspace}
\newcommand{\Sigmac}{\ensuremath{\Sigma_{c}^{0,++}}\xspace}
\newcommand{\Omegacz}{\ensuremath{\Omega_{c}^{0}}\xspace}

\let\Lc=\Lambdac

\newcommand{\DtoKpi}          {\ensuremath{D^{0} \to K^{-}\pi^{+}}\xspace}
\newcommand{\DplustoKKpi}     {\ensuremath{D^{+} \to K^{+} K^{-}\pi^{+}}\xspace}
\newcommand{\DstartoDzeropi}  {\ensuremath{D^{*+} \to D^{0}\pi^{+}}\xspace}
\newcommand{\Dstophip}        {\ensuremath{D_{s}^{+} \to \phi \pi^{+} \to K^{+}K^{-}\pi^{+}}\xspace}
\newcommand{\LctopKpi}        {\ensuremath{\Lambda_{c}^{+} \to pK^{-}\pi^{+}}\xspace}
\newcommand{\LctopKzeros}     {\ensuremath{\Lambda_{c}^{+} \to pK^{0}_{\rm S}}\xspace}

\usepackage{xspace} 
\usepackage{upgreek}

\def\MagUp {\mbox{\em Mag\kern -0.05em Up}\xspace}

{

 \def\PDelta      {\ensuremath{\Delta}\xspace}                 
 \def\PXi         {\ensuremath{\Xi}\xspace}                 
 \def\PLambda     {\ensuremath{\Lambda}\xspace}                 
 \def\PSigma      {\ensuremath{\Sigma}\xspace}                 
 \def\POmega      {\ensuremath{\Omega}\xspace}                 
 \def\PUpsilon    {\ensuremath{\Upsilon}\xspace}

 \def\PB      {\ensuremath{\mathrm{B}}\xspace}                 
                  
 \def\PD      {\ensuremath{\mathrm{D}}\xspace}

 \def\PK      {\ensuremath{\mathrm{K}}\xspace}

 \def\Pi      {\ensuremath{\mathrm{i}}\xspace}

 \def\thebaroffset{0.0em}
}
{

 \mathchardef\PDelta="7101
 \mathchardef\PXi="7104
 \mathchardef\PLambda="7103
 \mathchardef\PSigma="7106
 \mathchardef\POmega="710A
 \mathchardef\PUpsilon="7107
                  
 \def\PB      {\ensuremath{B}\xspace}                 
                  
 \def\PD      {\ensuremath{D}\xspace}

 \def\PK      {\ensuremath{K}\xspace}

 \def\Pi      {\ensuremath{i}\xspace}

 \def\thebaroffset{0.18em}
}
\newcommand{\offsetoverline}[2][\thebaroffset]{\kern #1\overline{\kern -#1 #2}}%

\def\to                 {\ensuremath{\rightarrow}\xspace}

\def\order   {{\ensuremath{\mathcal{O}}}\xspace}

\def\AT#1     {\ensuremath{A_{\mathrm{T}}^{#1}}\xspace}           % 2

\def\C#1      {\ensuremath{\mathcal{C}_{#1}}\xspace}                       % 9
\def\Cp#1     {\ensuremath{\mathcal{C}_{#1}^{'}}\xspace}                    % 7
\def\Ceff#1   {\ensuremath{\mathcal{C}_{#1}^{\mathrm{(eff)}}}\xspace}        % 9  
\def\Cpeff#1  {\ensuremath{\mathcal{C}_{#1}^{'\mathrm{(eff)}}}\xspace}       % 7
\def\Ope#1    {\ensuremath{\mathcal{O}_{#1}}\xspace}                       % 2
\def\Opep#1   {\ensuremath{\mathcal{O}_{#1}^{'}}\xspace}                    % 7

\newcommand{\aunit}[1]{\ensuremath{\text{\,#1}}}       
\newcommand{\kev}{\aunit{ke\kern -0.1em V}\xspace}
\newcommand{\ev}{\aunit{e\kern -0.1em V}\xspace}
 
\def\mbarn{\aunit{mb}\xspace}

\def\invnb {{nb}$^{-1}$\,}

\def\order{{\ensuremath{\mathcal{O}}}\xspace}

\def\gsim{{~\raise.15em\hbox{$>$}\kern-.85em
          \lower.35em\hbox{$\sim$}~}\xspace}
\def\lsim{{~\raise.15em\hbox{$<$}\kern-.85em
          \lower.35em\hbox{$\sim$}~}\xspace}

\def\sqs   {\ensuremath{\protect\sqrt{s}}\xspace}

\def\pt         {\ensuremath{p_{\mathrm{T}}}\xspace}

\def\pythia     {\mbox{\textsc{Pythia}}\xspace}

\def\tell1  {TELL1\xspace}
\def\ukl1   {UKL1\xspace}

\begin{document}
%\preprint{APS/123-QED}

\title{Open charm production cross section from combined LHC experiments in $pp$ collisions at $\sqrt{s} = 5.02$~TeV}
\author{Christian Bierlich\\
Dept. of Physics, Lund University, Sweden\\  \texttt{christian.bierlich@hep.lu.se}
  \and
Jeremy Wilkinson\\
GSI Helmholtzzentrum f\"{u}r Schwerionenforschung GmbH,\\ Darmstadt, Germany\\
\texttt{jeremy.wilkinson@cern.ch}
\and 
Jiayin Sun, Giulia Manca\\
Universit\`a degli studi di Cagliari \& I.N.F.N.,\\ Cagliari, Italy\\
\texttt{jiayin.sun@cern.ch}\\
\texttt{giulia.manca@cern.ch}
\and
Raphael Granier de Cassagnac\\
Laboratoire Leprince-Ringuet, CNRS/IN2P3, Ecole Polytechnique,\\
Institut Polytechnique de Paris, Palaiseau, France\\
\texttt{raphael@cern.ch}
\and
Jacek Otwinowski\\
The Henryk Niewodniczanski Institute of Nuclear Physics, \\ 
Polish Academy of Sciences, Cracow, Poland\\
\texttt{jacek.tomasz.otwinowski@cern.ch}
  }

\maketitle
\newpage
\begin{abstract}
Open charm production in proton-proton collisions represents an important tool to investigate some of the most fundamental aspects of Quantum Chromodynamics, from the partonic mechanisms of heavy-quark production to the process of heavy-quark hadronisation. Over the last decade, the measurement of the production cross sections of charm mesons and baryons in proton--proton (pp) collisions was at the centre of a wide experimental effort at the Large Hadron Collider. Thanks to the complementarity of the different experiments, the production of charm hadrons was measured over a wide transverse momentum region and in different rapidity ranges. In this paper, the measurements of the charm hadrons \Dzero, \Dstar, \Dplus, \Ds, \Lambdac and \Xicz performed by the ALICE, CMS and LHCb collaborations in pp collisions at the centre-of-mass energy $\sqrt{s}=\rm 5.02~\TeV$ are combined to determine the total charm-quark production cross section $\sigma_{c\bar{c}}$ in a novel data-driven approach. The resulting total \ccbar cross section is 
\begin{equation*}
\begin{aligned}
    \sigma_{\ccbar} (\mathrm{\pp},5.02\,\mathrm{\TeV}) ={} & %8340.9 \pm 217.8\mathrm{ (stat.) } 
    %{}^{+367.1}_{-367.2} \mathrm{ (syst.) } 
    %{}^{+362.5}_{-457.5} \mathrm{ (extr.) } \\ 
    %&{}^{+679.9}_{-0} (\Omega_c)~\mu\mathrm{b}.
    8.34 \pm 0.22\mathrm{ (stat.) } 
    {}^{+0.37}_{-0.37} \mathrm{ (syst.) } 
    {}^{+0.36}_{-0.46} \mathrm{ (extr.) } 
    {}^{+0.68}_{-0} (\Omega_c) \mbarn.
\end{aligned}
\end{equation*}

The measured charm hadron distributions and corresponding cross sections are compared with the most recent theoretical calculations.
\end{abstract}

\section{Introduction}
Measurements of charm hadron production at the Large Hadron Collider (LHC) represent a unique opportunity to study the properties of Quantum Chromodynamics (QCD) in hadron collisions~\cite{Andronic:2015wma}. Charm hadrons are produced in high-energy particle collisions through the fragmentation of charm quarks, which are generated in early hard-scattering processes. The most common theoretical approach to describe this production is based on the QCD factorisation theorem~\cite{Collins:1989gx}. In this framework, the production of hadrons containing charm or beauty quarks is calculated as a convolution of three independent terms: the parton distribution functions (PDFs) of the incoming hadrons, the cross sections of the partonic scatterings producing the heavy quarks, and the fragmentation functions (FF) that parametrise the non-perturbative evolution of a heavy quark into a given species of heavy-flavour hadron. Precise measurements of the production cross sections of the various charm mesons and baryons provide a unique way to establish the region of validity of the factorisation approach and to quantitatively study the properties of QCD in both the perturbative and non-perturbative regimes.

The LHC provides the ideal experimental apparatus to study charm production in \pp collisions. Starting from early 2010, LHC has collided protons at very high instantaneous luminosity ($\order(10^{33-34})\,$cm$^{-2}$s$^{-1}$) and at centre-of-mass energies ranging from $\sqrts=900$\,GeV to 13 TeV. In addition, the unique design and the complementarity of the four main LHC experiments, namely ALICE, ATLAS, CMS and LHCb, allow for measurements of the charm meson and baryon production cross sections as a function of transverse momentum (\pt) from zero up to hundreds of GeV/$c$, and from central to very forward rapidity~($y$). Over the course of the last decade, the four LHC experiments have measured the production cross sections of $\Dzero, \Dstar, \Dplus, \Ds$, $\Lambdac$, $\Xic$, $\Sigmac$ and $\Omegacz$ at $\sqrt{s}=2.76$~\cite{ALICE:2012inj}, 5.02~\cite{Aaij:2016jht,Sirunyan:2017xss,Sirunyan:2019fnc,Acharya:2019mgn, ALICE:2020wfu,ALICE:2020wla,ALICE:2021psx,ALICE:2021mgk,ALICE:2022ych}, 7~\cite{ALICE:2012mhc,ALICE:2012gkr,ATLAS:2015igt,ALICE:2017olh,Acharya:2017kfy} and 13~\TeV~\cite{Aaij:2015bpa,ALICE:2021bli,ALICE:2021rzj,CMS:2021lab,ALICE:2022cop}.

Perturbative calculations at next-to-leading order, with next-to-leading-log resummation~\cite{Kniehl:2005mk,Kniehl:2012ti,Cacciari:1998it,Cacciari:2012ny}, have proved to provide a successful description of the (strange and non-strange) charm mesons and their yield ratios, as a function of transverse momentum and rapidity~\cite{Andronic:2015wma,Aaij:2015bpa,Acharya:2019mgn}. However, theoretical calculations, where the charm fragmentation is tuned on \ee and $\mathrm{e^{+}p}$ measurements, are not able to provide a satisfactory description of the charm-baryon production cross sections at the LHC~\cite{Acharya:2017kfy, ALICE:2020wfu}. An alternative approach is offered by general purpose Monte Carlo event generators, such as \pythia\ \cite{Bierlich:2022pfr}. Here the charm production cross section is calculated to leading order, and corrected by a parton shower. Fragmentation is not handled by fragmentation functions, but rather by dynamical hadronisation models, which -- in particular with recent additions such as junction reconnection and rope hadronisation \cite{Christiansen:2015yqa,Bierlich:2014xba,Bierlich:2015rha,Bierlich:2022ned} -- provide a better description of the charm-baryon production cross sections. In this paper, we use \pythia 8.303, including relevant recent additions, to provide the shape of charm hadron distributions, in order to extrapolate to unmeasured regions.

 While total charm cross-section measurements exist at lower energies~\cite{Lourenco:2006vw,STAR:2012nbd,PHENIX:2010xji},  the total charm cross sections have been measured only in specific kinematic regions at the LHC.
 A recent phenomenological work uses the FONLL\cite{Cacciari:1998it, Cacciari:2012ny} framework to extrapolate the $\Dzero$ cross-section at $\sqrt{s}= 5.02~$\TeV measured by the ALICE and LHCb collaborations to the full phase space and derives an estimation of the total charm production cross-section\cite{Yang:2023kdq}. 
In this article, we review the existing measurements of the $\Dzero, \Dplus, \Ds$, $\Lambdac$, $\Xicz$ production cross sections as a function of \pt and $y$ performed by the ALICE, CMS and LHCb collaborations at $\sqrt{s}= 5.02~$\TeV~\cite{Aaij:2016jht,Sirunyan:2017xss,Sirunyan:2019fnc,Acharya:2019mgn, ALICE:2020wfu,ALICE:2020wla,ALICE:2021psx,ALICE:2021mgk,ALICE:2022ych} and combine them into a total charm cross section, extrapolated to the full phase space. 
The choice of the centre-of-mass energy was driven by the abundance of available measurements and by the need of providing a complete reference for existing and future heavy-ion measurements of charm production performed at the same nucleon-nucleon energy.
The available data on charm baryon production are used for the first time to provide an estimation of the total cross section. The extrapolation procedure is discussed in detail, including the assumptions made. 
The possible $1\%$ intrinsic charm contribution in the proton, whose existence has not been confirmed by experimental data, is expected to appear at large rapidity beyond the coverage of LHC experiments, and is not considered in the extrapolation.

The article is organised as follows. In Section~\ref{sec:Exp}, an overview of the different LHC experiments is presented, together with a description of the datasets used for the extrapolation. The extrapolation procedure is described in section \ref{sec:extrapolation-procedure}, and results are presented in section \ref{sec:results}.
\section{Experiments and data samples}
\label{sec:Exp}
\begin{figure}[!tbp]
\centering
\centering
\includegraphics[width=0.6\textwidth]{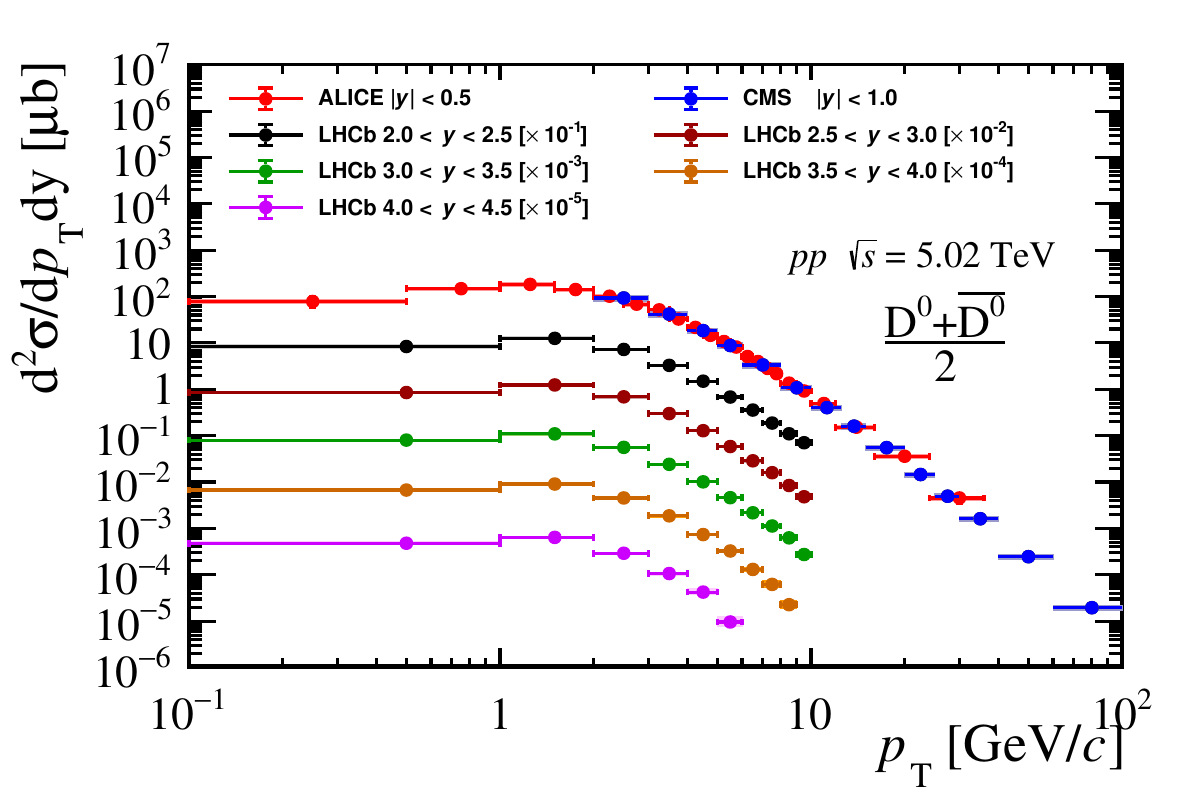}
\includegraphics[width=0.6\textwidth]{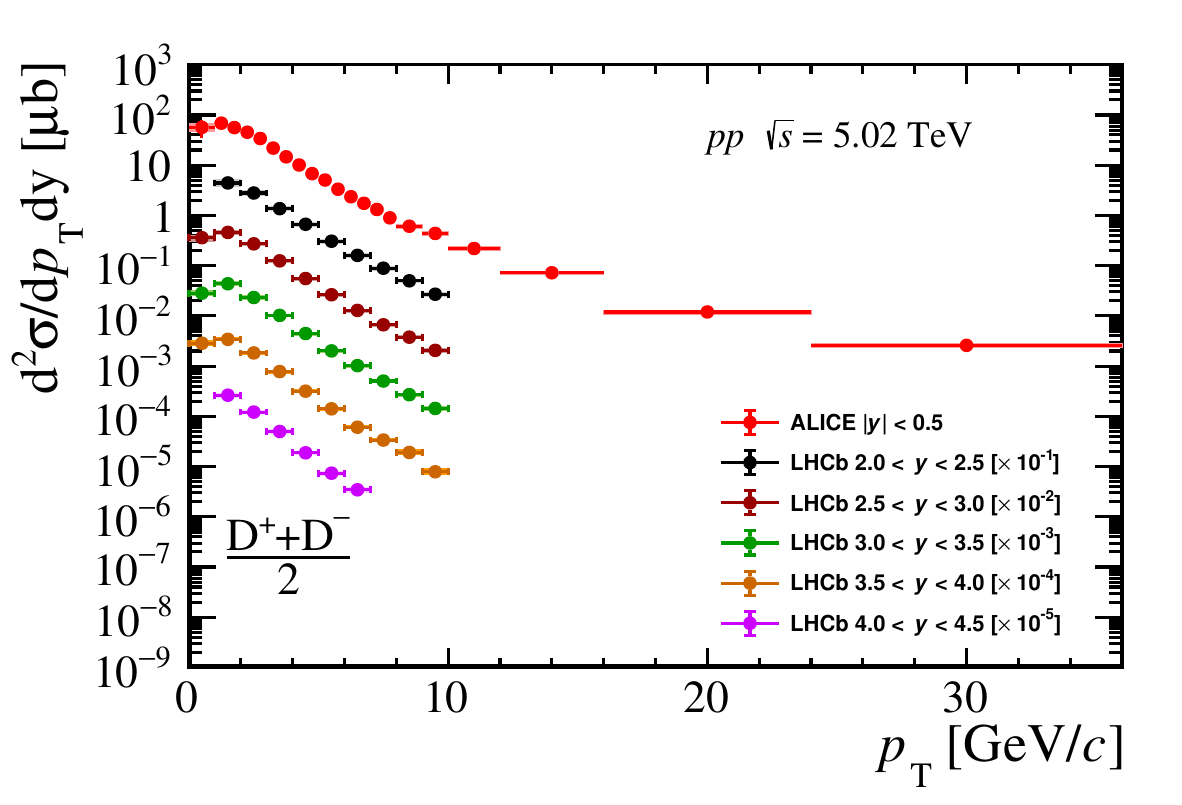}
\includegraphics[width=0.6\textwidth]{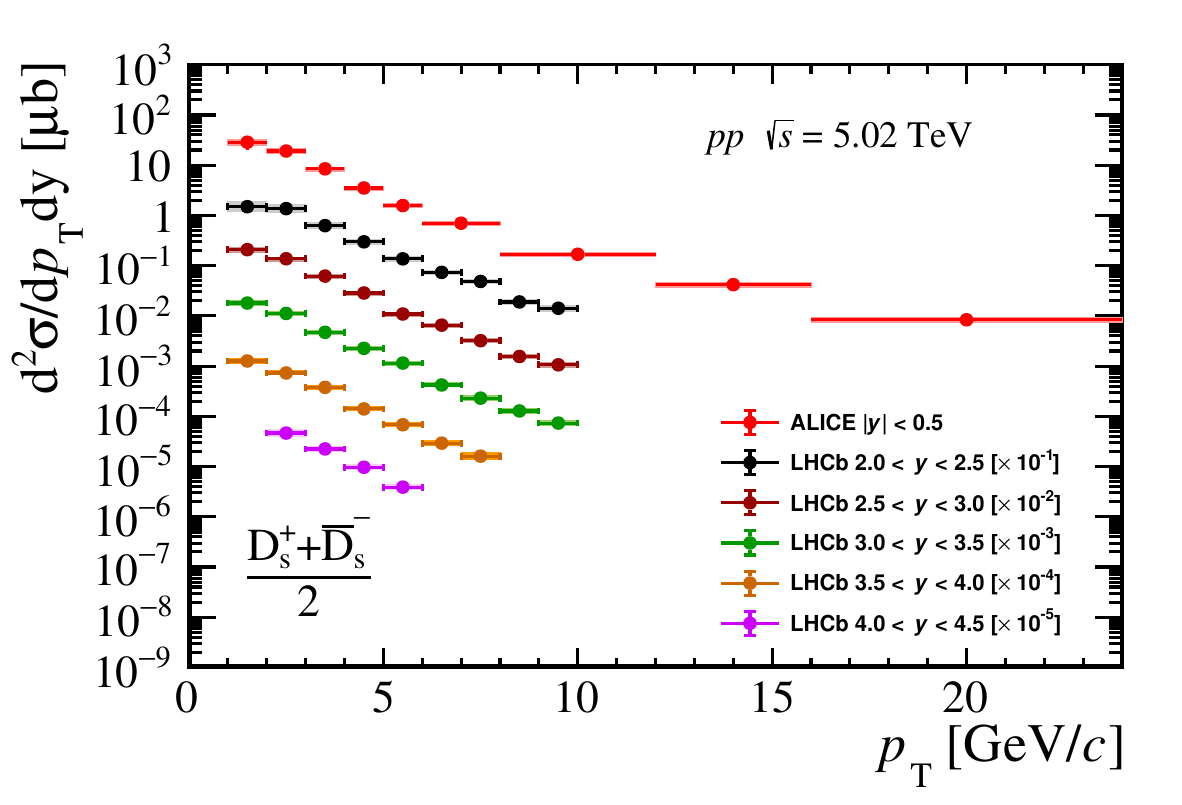}
\caption{Compilation of the measured double-differential cross sections of $\Dzero$ (top), $\Dplus$ (middle) and $\Ds$ (bottom) mesons as functions of \pt from the ALICE~\cite{Acharya:2019mgn, ALICE:2021mgk}, CMS~\cite{Sirunyan:2017xss} and LHCb~\cite{Aaij:2016jht} experiments.}
\label{fig:dmesons}
\end{figure}

The collision data used in this article have been delivered by the LHC operating at the CERN laboratory and
collected by the ALICE~\cite{Aamodt:2008zz},
CMS~\cite{Chatrchyan:2008aa} and LHCb~\cite{Alves:2008zz}
collaborations. The results are based on recent measurements published by the ALICE~\cite{Acharya:2019mgn, ALICE:2020wfu,ALICE:2020wla,ALICE:2021psx,ALICE:2021mgk,ALICE:2022ych},
CMS~\cite{Sirunyan:2017xss,Sirunyan:2019fnc} and LHCb~\cite{Aaij:2016jht} collaborations on the open charm production. A summary of the results used is given in
Tab.~\ref{tab:inputAna}, and an 
overview consisting of $D$-meson cross sections as function of \pt, is reported in Fig.~\ref{fig:dmesons}. The integrated luminosities of the data samples are 19.3~nb$^{-1}$, 27.4~pb$^{-1}$ and 8.6~pb$^{-1}$ for the
ALICE, CMS, and LHCb analyses, respectively.

\begin{table}
  \centering
  \begin{tabular}{cccc}
    \hline
    {\bf{Experiment}} & {\bf ALICE} & {\bf CMS} & {\bf LHCb} \\\hline
    Luminosity (pb$^{-1}$)  & (19.3$\pm$0.4)$\times 10^{-3}$ & 27.4$\pm$0.6  & 8.60$\pm$0.33\\
    Hadrons              & \Dzero, \Dplus, \Dstar, \Ds, $\Lambdac$, $\Xicz$ & \Dzero, $\Lambdac$ & \Dzero, \Dplus, \Dstar, \Ds \\
    \pt coverage (\gevc)   & 0--36 & 2--100 & 0--10 \\
    $y$ coverage           & $|y|<$0.5 & $|y|<$1.0 & 2.0$<y<$4.5 \\
    \hline
    \end{tabular}
    \caption{\label{tab:inputAna} The summary of the considered data sets.}
\end{table}

In the following, we briefly summarise the methodology used by each experiment in their own publications. 
\subsection{ALICE}
The ALICE collaboration measured the cross section of prompt $\Dzero, \Dplus, \Dstar, \Ds$, $\Lambdac$, and \Xicz hadrons and their respective charge conjugates in the range of rapidity $|y|<$~0.5, 
using a data sample of about 990 million Minimum Bias (MB) events, which were collected during the 2017 pp run. This corresponds to an integrated luminosity of about 19.3~\invnb. The charm hadrons were reconstructed via the hadronic decay channels \DtoKpi, \DplustoKKpi, \DstartoDzeropi with \DtoKpi, \Dstophip, \LctopKpi, and \LctopKzeros$\to p\pi^+\pi^-$, and the semileptonic decay channel $\Xicz\to\mathrm{e}^+\Xi^-\nu_\mathrm{e}$~\cite{Workman:2022ynf}. 

The prompt candidates are separated from the non-prompt coming from b-hadron decays either using predictions from FONLL, or by a data-driven approach based on the impact parameter distributions of the candidates. A different analysis technique is applied for the prompt $\Dzero$ reconstruction, which is mainly based on particle identification and combinatorial background subtraction. It allows for extending $\Dzero$ measurement down to $\pt=0$.

The \pt-differential cross sections of prompt hadrons were obtained as
\begin{equation}
\left.\frac{{\rm d}\sigma}{{\rm    d}\pt}\right|_{\scriptscriptstyle |y_{\rm lab}|<0.5}=\frac{f_{\rm{prompt}}\cdot \left.N_{\rm raw}\right|_{|y_{\rm lab}|<y_{\rm \scriptscriptstyle fid}}}{2 \cdot \alpha_{\rm \scriptscriptstyle y} \cdot \Delta\pt \cdot ({\rm Acc} \times
   \epsilon)_{{\rm prompt}}\cdot {\rm{BR}}\cdot \mathcal{L}}.
\end{equation}

Here, $N_{\rm raw}$ is the raw yield extracted in a given $\pt$ interval (of width $\Delta\pt$), $({\rm Acc}\times\epsilon)_{{\rm prompt}}$ is the geometrical acceptance multiplied by the reconstruction and selection efficiency of prompt hadrons, and $f_{{\rm prompt}}$ is the fraction of prompt hadrons in the uncorrected yield. The factor $\alpha_{\rm \scriptscriptstyle y}$ accounts for the \pt-dependent fiducial acceptance selection on the candidates. A factor $2$ was added to the denominator to take into account that anti-particles are counted in the raw yield, but the corrected yields are given for  only one particle type. 
Finally, $\rm{\mathcal{L}} = \rm N^{MB}_{ev}/\sigma_{MB}$, where $\rm N^{MB}_{ev}$ is the number of analysed MB events and $\rm \sigma_{MB}=$ 50.9~mb is the inelastic cross section for the MB trigger condition.

A detailed description of the reconstruction and selection of the various charm hadrons, including the 
efficiency estimation and the systematic uncertainty evaluation, can be found in Refs.~\cite{Acharya:2019mgn,ALICE:2021psx,ALICE:2021mgk,ALICE:2022ych}.

\subsection{CMS}
The CMS collaboration measured charm cross sections for both prompt \cite{Sirunyan:2017xss} and non-prompt \cite{Sirunyan:2018ktu} $\Dzero$ production in both pp and heavy-ion collisions through the $\DtoKpi$ decay.
The prompt cross section for pp collisions uses data samples with luminosity up to 27.4 pb$^{-1}$
from the run at $\sqrt{s}=5.02$ TeV recorded in 2015.
The $\Dzero$ rapidity was limited to $|y|<1$ in the analysis to 
profit from the best possible tracking resolution, and from a dedicated trigger for the high-\pt part of the measurement which allowed the extension of the measurement up to
$\pt=100$~\gevc \cite{Sirunyan:2017xss}. The high background due to the absence of dedicated pion and kaon identification limited the lower momentum of the measurement to transverse momenta above 2~\gevc, in a sample of highly prescaled MB triggered events. The cross sections originally quoted 
for the sum of $\Dzero$ and ${\overline{D}{}^0}$ were divided by 2 for the purpose of this work, in order to make them consistent with the ALICE convention. 

In a separate result \cite{Sirunyan:2019fnc}, 
\LctopKpi final states were measured in the same rapidity range $|y|<1$ on the same data set.
This measurement was limited to the transverse momentum region $5<p_T<20$~\gevc.
 
The CMS results overlap largely with the ALICE measurements which exhibit higher precision at low~\pt. Since the total cross section is driven by the lower \pt spectrum, the CMS measurements  are not included in the final combination but used for systematic studies.

\subsection{LHCb}
The LHCb collaboration measured the cross section of prompt $\Dzero$, $\Dplus$, $\Ds$ and $\Dstar$ mesons at centre-of-mass energy
$\sqrt{s}=5.02$~TeV, using a data sample with an integrated luminosity of $8.60\pm0.33$ pb$^{-1}$ recorded in 2015. The prompt component is separated from the non-prompt coming from b-hadron decays by using the high resolution of the vertex locator (VELO) detector.  Thanks to the unique LHCb
coverage in the forward rapidity region, the measurement is performed in the $\pt$ range of $0<\pt<10$~\gevc for $\Dzero$ and $\Dplus$ mesons, and $1<\pt<10$~\gevc for $\Ds$ and $\Dstar$ mesons. The measured rapidity range covers the region \nobreak{$2.0<y<4.5$}. The final states 
$\DtoKpi$, $D^+ \rightarrow K^- \pi^+ \pi^+$, \Dstophip and $\Dstar \rightarrow (D^0\rightarrow K^- \pi^+) \pi^+$ and their charge conjugated are reconstructed. A detailed description of the reconstruction, selection of signals and efficiency determination is discussed in Ref.~\cite{Aaij:2016jht}.
The double-differential cross section is reported as a function of \pt and $y$:
\begin{equation}
    \frac{\mathrm{d}^2\sigma}{\mathrm{d}p_\mathrm{T}\mathrm{d}y}=\frac{1}{\Delta p_\mathrm{T}\Delta y}\times\frac{N_\mathrm{D}}{\epsilon_D \times \rm{BR} \times \ensuremath{\mathcal{L}}}
\end{equation}
where $\Delta p_\mathrm{T}=1$~\gevc and $\Delta y=0.5$ are the widths of $p_\mathrm{T}$ and $y$ bins, $N_\mathrm{D}$ is the measured signal yield of the D-meson plus the charge-conjugated yields, $\epsilon_D$ is the total efficiency for the D-meson, $\rm{BR}$ is the branching ratio of the decay and $\mathcal{L}$ is the total integrated luminosity.

\section{Extrapolation procedure}
\label{sec:extrapolation-procedure}
As stated in the introduction, the ultimate goal of this paper is to obtain the total $\ccbar$ cross section by extrapolating LHC measurements using \pythia\ for the estimation in the unmeasured regions of phase space. 
In this section we will first provide details on the simulation by introducing the relevant parts of \pythia\ in section \ref{sec:pythia}. We then proceed explaining the estimation of the only remaining free parameter, namely the kinematic charm mass, in section \ref{sec:kinematic-charm-mass}. Finally, in section \ref{sec:ext-procedure} we explain the extrapolation procedure itself, with an estimation of extrapolation uncertainty in section \ref{sec:alternative}.

\subsection{PYTHIA}\label{sec:pythia}

The Monte Carlo event generator \pythia \cite{Bierlich:2022pfr} is one of the standard tools for generating simulated collision events at most of the collider experiments. In this section, the main model ingredients relevant for charm-hadron production are introduced.

When simulating charmed hadron production down to very low \pt, both perturbative and non-perturbative aspects need to be considered \cite{Norrbin:2000zc}. At leading order (LO),  $\mathcal{O}(\alpha^2_s)$, the processes $q\bar{q} \rightarrow c\bar{c}$ and $gg \rightarrow c\bar{c}$ define the starting point in \pythia. A possible, and common, evolution from the LO processes is to add modifications from new processes in increasing orders of $\alpha_s$. The alternative, which is used by \pythia, is the parton-shower approach. While not exact, even to $\mathcal{O}(\alpha^3_s)$, it is correct to leading logarithmic order (LL). This approach adds the possibility of gluon splitting ($g \rightarrow c\bar{c}$) in the shower, as well as flavour excitation, where a charm quark from the parton distribution function is put on mass shell by an interaction, allowing charm quarks to be produced in the shower down to very low \pt.

The hard $2\rightarrow2$ production requires a suitable, process-dependent phase space cut to be introduced, to avoid soft and collinear singularities. In this study, we use the \pythia model for multiparton interactions \cite{Sjostrand:1987su}, which introduces an effective parameter $p_\mathrm{T,0}$ based on arguments of colour screening effects. In this way we allow for a smooth regularisation of the cross section, and, in combination with the parton shower approach outlined above, production of charm quark pairs down to non-perturbative scales.

As the $p_\mathrm{T,0}$ parameter is fixed by the total multiplicity and average transverse momentum $\langle \pt \rangle$ of charged particles \cite{Skands:2014pea}, the only remaining parameter to fix is the charm mass. In \pythia, the default value is 1.5 GeV/$c^2$, based on previous fits to low-energy data of D-meson production \cite{Norrbin:1998bw}.

At this point in the event generation process, the quarks and gluons produced from hard interactions and parton shower, must be transformed to hadrons. In \pythia this is handled by the Lund string fragmentation model \cite{Andersson:1983ia}.
When quarks and gluons move apart, a colour flux tube modelled as a massless relativistic string with tension $\kappa \approx 1$ GeV/fm is stretched between them. Once they are far enough apart that it is energetically favourable to the string to break into hadrons, it will do so, producing new quark-anti-quark or diquark-anti-diquark pairs in the string breaking.
Charm quarks are too heavy to be produced in string breakings. This means that once the amount of charm quarks available for production of heavy-flavour hadrons is decided by the perturbative calculation, the total charm production is fixed. 
The simplest model for baryon production lets the string break into either quarks or diquarks, with diquark breakings resulting in baryon production. Baryons can, however, also be produced by another mechanism, through the creation of string junctions in the so-called colour reconnection stage. The main motivation for introducing such models is explained in the following. The parton shower introduced above is derived in the leading-colour ($N_c \rightarrow \infty$) approximation, which means that colour flow configurations are uniquely defined by the shower, and no coherence effects between strings exist. Several models to correct this approximation on the nonperturbative level exist, and while primarily intended to explain the rise of multi-strange baryons with multiplicity \cite{Bierlich:2014xba,Christiansen:2015yqa,ALICE:2017jyt} and flow signatures in proton collisions \cite{Bierlich:2017vhg,Khachatryan:2010gv}, they also have consequences in the charm sector.
The models relevant for this paper are concerned with regions in real space densely populated with strings. When the strings overlap, coherence effects must be accounted for. Simple $\mathrm{q}\bar{\mathrm{q}}$ strings are $SU(3)$ triplets. When they overlap with each other, the resulting structure is in a higher multiplet state, some of which are lower multiplets associated with string junctions \cite{Christiansen:2015yqa,Sjostrand:2002ip}, and the highest multiplet, a so-called ``rope" \cite{Bierlich:2014xba,Biro:1984cf}. The junctions have colour flows connecting multiple quarks in three-quark vertices. As such, they carry an intrinsic baryon (or anti-baryon) number, and will hadronise as baryons. The highest multiplet has a higher string tension than a normal triplet string, and when it hadronises, the overall suppression of \textit{e.g.} strange quark production, will decrease. In the case of charm production it means, to first order, that relatively more $\Lambdac$ baryons will be produced, at the expense of D-mesons, and that relatively more $\Ds$ mesons will be produced, at the expense of $\Dplus$. Since this analysis uses \pythia to extrapolate measured spectra of hadrons to unmeasured regions, it is important to consider these effects.
All the models introduce several new parameters, all of which are estimated in the light-flavour sector. 
Parameters which have previously been shown to provide a good description of the multi-strange baryons in the light sector, are used here as well, and are summarised in Tab.~\ref{table:pythia-parameters} in Appendix~\ref{appendix:parameters}.

For this study, the most relevant parameter available for tuning is the charm quark mass, as long as the agreement with other data sets is not compromised, as will be explained in the following.

\subsection{Estimating the kinematic charm mass}
\label{sec:kinematic-charm-mass}

As explained in section \ref{sec:pythia}, the only remaining parameter left influencing the cross section of charmed hadrons is the charm quark mass. The charm mass enters in both
the perturbative matrix elements and in the phase space selection, and the resulting \textit{kinematic charm mass} is therefore not required to
be identical to the current quark mass, though it should not deviate too far from it. In ref. \cite{Norrbin:1998bw} a value of
$m_{c}=1.5$ GeV/$c^2$ was chosen, loosely based on data from WA82 \cite{WA82:1993ghz}, E769 \cite{E769:1993hmj}, and E791 \cite{E791:1996htn}. We updated this value using the large amount of recent data on D-meson cross sections from the LHC (Fig.~\ref{fig:dmesons}).

\begin{figure}[!tbp]
\centering
\begin{minipage}[t]{0.80\textwidth}
\centering
\includegraphics[width=1.0\textwidth]{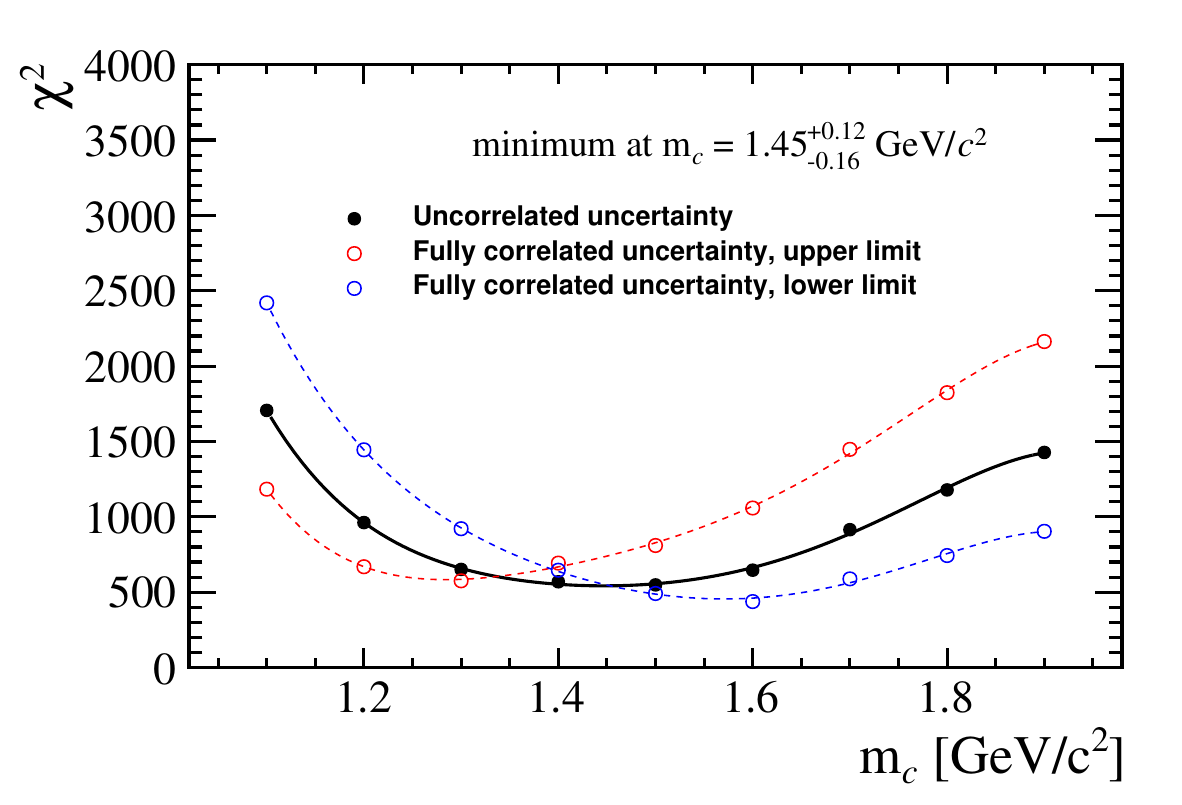}
\end{minipage}
	\caption{Fit of the kinematic charm mass. The global $\chi^2$ between \pythia and all available $\mathrm{D}$-meson data, fitted with a fifth order polynomial, is shown.}
\label{fig:chi2fit}
\end{figure}

In \pythia simulations, a variation of the charm quark mass $m_{c}$ is performed from $m_{c}=1.1$ GeV/$c^2$ to $1.9$ GeV/$c^2$ with steps of $\Delta m_{c}=0.1$~GeV/$c^2$ (see Figs.~\ref{fig:pythia_mc_d0}, \ref{fig:pythia_mc_dp} and \ref{fig:pythia_mc_ds} in Appendix~\ref{appendix:plots}). The $\chi^2$ values are computed between the data and \pythia simulations for each $m_{c}$ value. All available $\Dzero$, $\Dplus$ and $\Ds$ data points with $\pt <6\,\GeVc$ are used in the calculation to reduce statistical fluctuations in the data and \pythia simulation at higher \pt, yielding a total of 4 data points from CMS data, 28 from ALICE and 82 from LHCb. For the systematic uncertainties in these data points, the uncertainties between experiments are uncorrelated except for those from the branching ratios. Within each experiment, the CMS points are assumed to be uncorrelated, while both the ALICE and the LHCb uncertainties are partially correlated. As the correlation between systematic uncertainties in these measurements is not available, two extreme scenarios are considered: a) all experimental uncertainties between data points are uncorrelated, b) all systematic uncertainties between ALICE and LHCb data points are fully correlated. 
Hence all the data points are shifted up and down by $1\sigma$ from their nominal values, and the corresponding $\chi^2$ values are calculated. The resulting $\chi^2$ versus the charm quark mass $m_{c}$ is shown in Fig.~\ref{fig:chi2fit}, where the black points denote the uncorrelated scenario, and the red and blue show the fully correlated scenario with the upper and lower limit, respectively. Fifth-order polynomial functions are used to fit the kinematic charm mass. The minimum $\chi^2$ corresponds to the $m_{c}$ value that agrees best with the data, which is found to be $m_{c} = $ 1.29,  1.45, and 1.57 $\mathrm{GeV}/c^2$ for the red, black and blue points, respectively. The differences between these $m_{c}$ values are considered as the systematic uncertainty originated from the data uncertainty. 
The minimum $\chi^2$ values around 500 indicate that \pythia does not reproduce the D-meson data perfectly.

\begin{figure}[!tbp]
\centering
\begin{minipage}[t]{1.00\textwidth}
\centering
\includegraphics[width=1.0\textwidth]{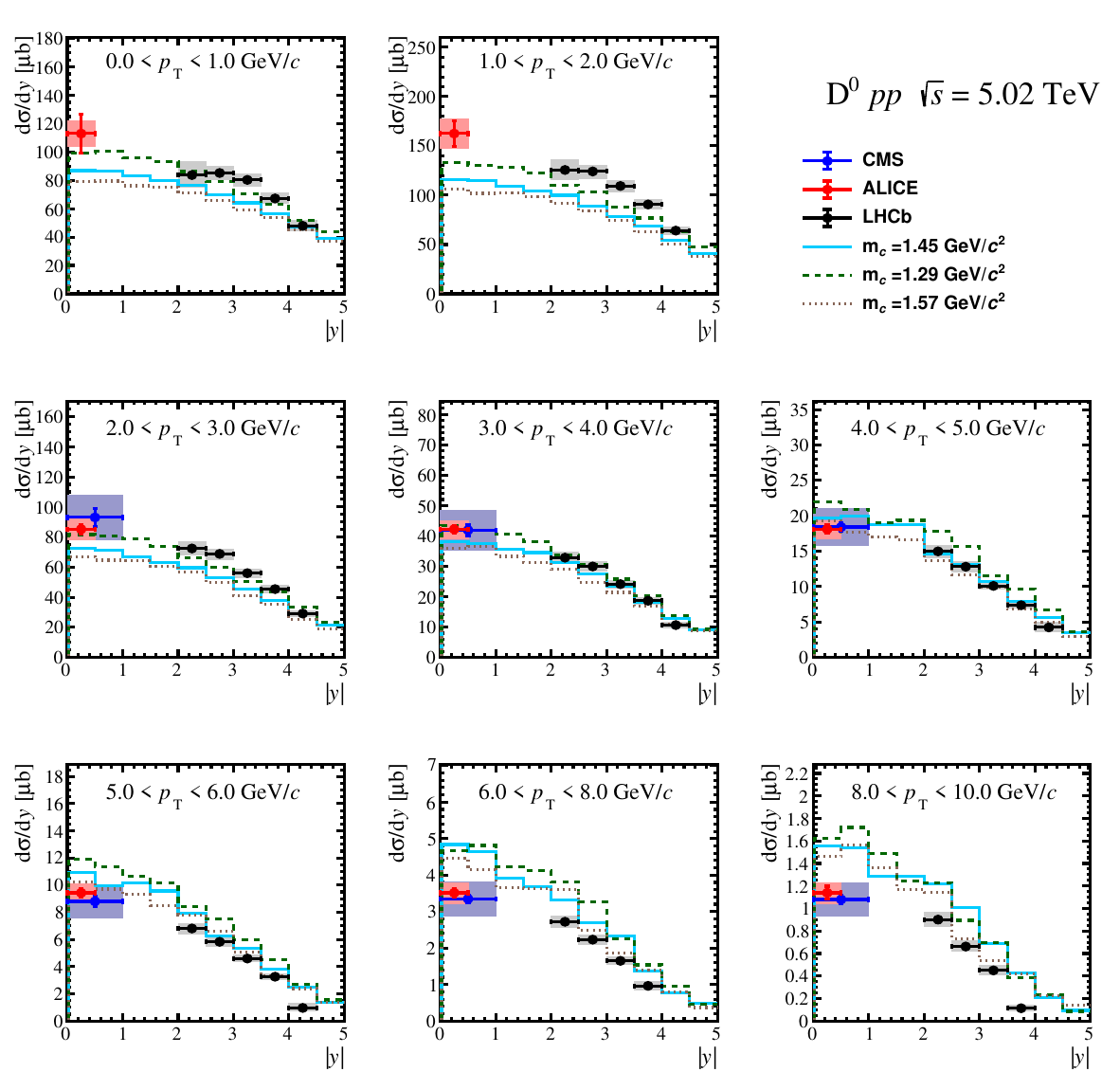}
\end{minipage}
	\caption{Measured $\Dzero$ cross section as a function of rapidity compared with \pythia simulations with the bands corresponding to a 1$\sigma$ variation around the optimum.
	}
\label{fig:pythia_mc_d0_result}
\end{figure}

\begin{figure}[!tbp]
\centering
\begin{minipage}[t]{1.00\textwidth}
\centering
\includegraphics[width=1.0\textwidth]{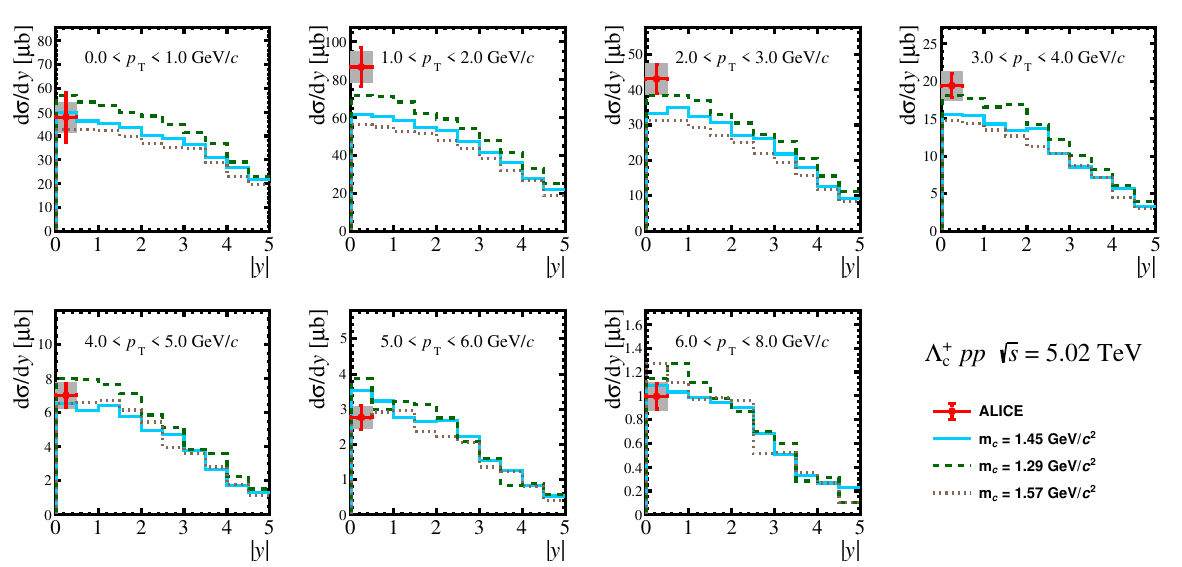}
\end{minipage}
	\caption{Measured $\Lambdac$ cross section as a function of rapidity compared with \pythia simulations with the bands corresponding to a 1$\sigma$ variation around the optimum.
	}
\label{fig:pythia_mc_Lc_result}
\end{figure}

The resulting value ($\pm 1\sigma$) for the kinematic charm mass, which will be used in the following, is:
\begin{equation}
	m_{c} = 1.45^{+0.12}_{-0.16} \mathrm{GeV}/c^2.
\end{equation}
 Figures \ref{fig:pythia_mc_d0_result} and \ref{fig:pythia_mc_Lc_result} show the measured cross sections for \Dzero and \Lambdac compared to the \pythia\ simulations with the bands corresponding to $1\sigma$ variations around the kinematic charm mass. The same comparison is shown in appendix~\ref{appendix:plots} for \Dplus (Fig.~\ref{fig:pythia_mc_dp_result}), \Ds (Fig.~\ref{fig:pythia_mc_ds_result}) and \Xicz (Fig.~\ref{fig:pythia_mc_xc_result}). 
 
 We recommend to use this value of kinematic charm mass in future charm hadron studies at LHC, along with the model parameters given in Tab.~\ref{table:pythia-parameters} in App.~\ref{appendix:parameters}.

\clearpage 

\subsection{Extrapolation to total $\ccbar$ cross section}
\label{sec:ext-procedure}

The total charm cross section in pp collisions at \sqrtsfive is derived by extrapolating the available \pt- and $y$-differential cross sections of charm hadrons to the range $0<\pt<36\,\GeVc$ and $|y|<8$. The contribution to the total cross section above these bounds is considered to be negligible with respect to that at lower \pt, as the $\mathrm{d}\sigma/\mathrm{d}y$ value for charm from FONLL pQCD calculations \cite{Cacciari:1998it} falls to 0 for $|y|<8$, and similarly less than $0.01\%$ of the total \pt-differential cross section lies above $\pt=36\, \GeVc$. 

The measured charm-hadron production cross sections are integrated in their visible ranges with a consistent treatment of correlations in their systematic uncertainties as described in their respective publications. When combining the measurements together, the uncertainties are assumed to be uncorrelated between experiments, apart from the branching ratio uncertainty, which is treated as fully correlated.

The extrapolation is performed separately for each hadron species, and in two steps, assuming that the \pt and $y$ dependences are factorisable. First, an extrapolation in \pt (where needed) is performed within each experiment's rapidity range ($|y| < 0.5$ for ALICE; $2.0<y<4.5$ for LHCb). The extrapolation factor is determined by dividing the integrated \pythia\ spectrum for $\pt < 36\,\GeVc$ by that within the visible \pt range. Then it is multiplied by the integrated visible production cross section. The statistical and systematic uncertainties on the integrated measurements are scaled by the same factor. The respective \pt extrapolation factors for each hadron species and experiment are detailed in Tab.~\ref{tab:ptextrapfacs}.

The extrapolation in rapidity is performed using a similar method, by taking the ratio of the integrated \pythia\ spectra in $|y|< 8$ over the visible rapidity range for $\pt<36$\,\GeVc. For the D-meson species, where measurements are available for both ALICE and LHCb, the visible range comprises the rapidity acceptances of both detectors together; for the charm baryon species where only ALICE measurements exist, the visible range was taken as $|y|<0.5$. The rapidity interpolation factors are detailed in Tab.~\ref{tab:rapinterpfacs}.

The total charm production cross section, $\sigma_{\ccbar}$, is calculated from the measured ground-state hadron species as

\begin{equation}
    \sigma_{\ccbar} = \sigma(\Dzero)+\sigma(\Dplus) + \sigma(\Ds) + \sigma(\Lc) + 2 \times \sigma(\Xicz).
    \label{eq:combin}
\end{equation}

As no measured cross section is available for $\Xicp$ at this collision energy, a factor 2 is applied on the $\Xicz$ cross section in the calculation. This is motivated by the assumption that due to isospin symmetry the production rates of $\Xicp$ and $\Xicz$ are equivalent, and so enter into the total charm production cross section equally. At very forward rapidity, the possibility of an enhanced production of $\Xicp$ over $\Xicz$ is not considered due to lack of experimental data. The contribution of the $\Omega_\mathrm{c}$ baryon is considered to be small with respect to the other hadron species and so is not added to the central value. However, to account for the possible case that the $\Omega_\mathrm{c}$ enters significantly into the total charm cross section, an upper systematic uncertainty is assigned based on the $\Xic$ cross section, under the extreme assumption of $\Omega_\mathrm{c}/\Xicz=1$.

An extrapolation uncertainty is assigned to the total cross section by varying the bare charm mass used as an input parameter of \pythia\ as discussed above and recalculating the central value of the extrapolated cross section, resulting in an extrapolation uncertainty of approximately 2\%. As the rapidity dependence of $\Xic$-baryon production is not well studied yet in pp collisions, an additional version of the extrapolation was performed, where the Monash tune of \pythia\ was used instead of the enhanced colour reconnection (CR) mode discussed above. This tune of \pythia\ predicts a significantly smaller charm baryon-to-meson ratio than the CR mode 2 for both \Lc and \Xic baryons as measured by the \mbox{ALICE} collaboration at mid-rapidity. After extrapolating with this model, the relative contribution of \Xic baryons is reduced by approximately 7\%, with corresponding increases in the relative abundances of \Dzero, \Dplus, and \Lc. The resulting total charm production cross section is lower by 8\% with respect to the central prediction; this value is assigned as a lower bound on the extrapolation uncertainty in the final result.

\subsection{Alternative extrapolation}
\label{sec:alternative}
An alternative, model-independent extrapolation method is employed to cross-check the result by replacing the PYTHIA spectrum with numerical functions fitted to data. The two step extrapolation procedure remains unchanged. First, in each visible $y$ interval, a Tsallis function~\cite{Tsallis} is used to fit the measured $\pt$-differential cross sections. The extrapolation factor is determined as the ratio of the integral of the Tsallis function in the range $0<\pt<36$ \GeVc to that within the visible \pt range. The integrated visible cross section is then multiplied by the factor. This step is performed for \Dzero, \Dplus and \Ds mesons in the $y$ intervals of $0<y<0.5$, $2<y<2.5$, $2.5<y<3$, $3<y<3.5$, $3.5<y<4$ and $4<y<4.5$, and for \Lc and \Xicz baryons in $0<y<0.5$. 
Next, for the extrapolation in rapidity, a Gaussian function is used to fit the extrapolated cross section in $0<\pt<36$ \GeVc as a function of $y$ for each D-meson species. The extrapolation factor in rapidity is determined similarly, using $|y|<8$ as the full rapidity range and the sum of the ALICE and LHCb acceptance as the visible region. The final extrapolated total cross section is compared to that extrapolated with PYTHIA, showing a $\sim 4\%$ variation in the final value. This variation is added in quadrature as an extra contribution to the extrapolation uncertainties in the result.
\section{Results and comparison to models}
\label{sec:results}

The data from the various experiments compared with \pythia\ with the dedicated tuning discussed in Sec.~\ref{sec:pythia} are reported in Figs. ~\ref{fig:pythia_mc_d0_result}, \ref{fig:pythia_mc_Lc_result}, \ref{fig:pythia_mc_dp_result}, \ref{fig:pythia_mc_ds_result},  and \ref{fig:pythia_mc_xc_result} as a function of rapidity and \pt.

The total charm production cross section in pp collisions at $\sqrt{s}=5.02~\mathrm{TeV}$, as obtained from Eq.~\ref{eq:combin}, is
\begin{equation}
    %\sigma_{\ccbar} = 8340.9 \pm 217.8\mathrm{ (stat.) } {}^{+367.1}_{-367.2} \mathrm{ (syst.) } {}^{+362.5}_{-457.5} \mathrm{ (extr.) } {}^{+679.9}_{-0} (\Omega_c)~\mu\mathrm{b}.
    \sigma_{\ccbar} = 8.34 \pm 0.22\mathrm{ (stat.) } {}^{+0.37}_{-0.37} \mathrm{ (syst.) } {}^{+0.36}_{-0.46} \mathrm{ (extr.) } {}^{+0.68}_{-0} (\Omega_c) \mbarn .%/times~m\mathrm{b}.
\end{equation}

\begin{figure}[t]
\centering
\includegraphics[width=0.95\textwidth]{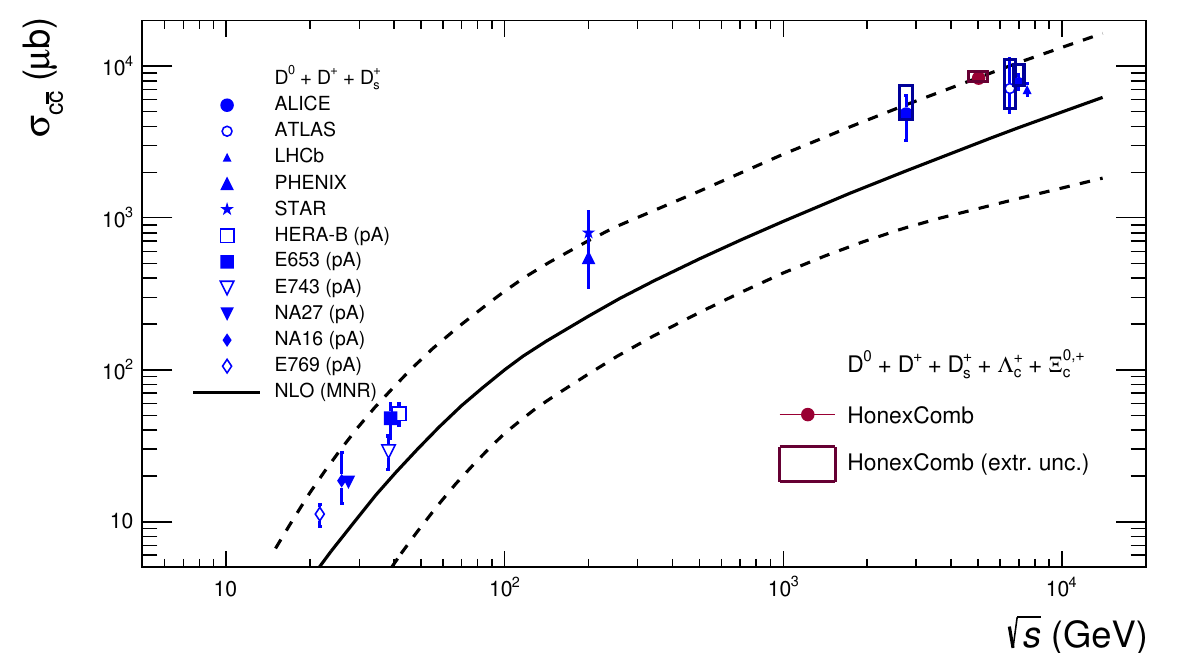}
\caption{The total charm production cross section measured in p--A and pp collisions by experiments~\cite{Lourenco:2006vw,STAR:2012nbd,PHENIX:2010xji} as a function of collision energy. For the measurement reported from this paper, the experimental uncertainties are shown as error bars and the extrapolation-related uncertainties are shown as a box. The measurements are compared with the NLO (MNR) predictions represented by the solid line \cite{Mangano:1991jk}. The dashed lines denote the systematic uncertainties on the calculations (see text for more details).}
\label{fig:totalcharmxsec}
\end{figure}

The uncertainty sources are as described in Sec.~\ref{sec:extrapolation-procedure}. The value is also compared with model calculations from the next-to-leading order MNR pQCD framework~\cite{Mangano:1991jk} in Fig.~\ref{fig:totalcharmxsec} and with the charm cross sections measured by other experiments in pp and p--A collisions at different centre-of-mass energies \cite{Lourenco:2006vw,STAR:2012nbd,PHENIX:2010xji}. It should be noted that the other experimental points shown, in particular those reported by the LHC collaborations at $\sqrt{s}=2.76\,\mathrm{TeV}$ and $7\,\mathrm{TeV}$, consider only the contributions of $D$-mesons scaled by the estimated fragmentation fractions from $\rm e^+e^-$ collisions, and do not include measurements of charm-baryon production.

The uncertainties on the NLO (MNR) calculation are defined by summing in quadrature the contributions from a) variations in the factorisation and renormalisation scales $\mu_\mathrm{f}$ and $\mu_\mathrm{r}$ between 0.5 and 2, with the constraint $0.5 < \mu_\mathrm{f}/\mu_\mathrm{r} < 2$; b) variations in the charm mass between $m_c=1.2$ and $1.8\,$\GeVc; and c) the uncertainties related to the applied PDF set (CTEQ6.6~\cite{Nadolsky:2008zw}). The  value of the $\ccbar$ cross section is more than double the central value of the predictions, but is fully consistent within the theoretical uncertainties. It should be noted that the central value of the NLO calculations assumes a larger charm mass $(m_c = 1.5\,\mathrm{GeV}/c)$ than the one determined above from fits to the experimental data, which strongly contributes to the lower predicted central value of the cross section.

\section{Conclusions}
The total $\ccbar$ cross section has been extrapolated using the open-charm-hadron production measurements from the three LHC experiments ALICE, CMS and LHCb collected from pp collisions at $\sqs=5.02$~\tev.
The results have been presented as a function of the rapidity of the charm mesons and have, for the case of differential results, been compared to \pythia, including recent improvements in hadronisation models, and for the case of the integrated result, with a state-of-the-art fixed order calculation.

The main result, shown in Fig.~\ref{fig:totalcharmxsec}, represents the most comprehensive measurement to-date of the total \ccbar cross section in pp collsions. We note that along with an increase in precision with respect to the earlier results due both to the increase in experimental precision for the LHC experiments between Run 1 and 2 of the LHC and the reduced reliance on model-dependent extrapolations in rapidity, the measured $\ccbar$ cross section is now exactly at the edge of the uncertainty estimate of the NLO calculation.

The extrapolation procedure has been extensively described, being based on extracted shapes from \pythia, followed by a data driven approach.
We believe that this novel extrapolation procedure, and the slight update in the values of the parameter used in the shapes extraction,
can be adopted in the future for this type of extrapolation at different centre-of-mass energies, as the ones previously and currently being measured at the LHC as well as at other accelerators.

\newenvironment{acknowledgement}{\relax}{\relax}
\begin{acknowledgement}
\section*{Acknowledgements}
We thank the LHC collaborations for providing data in an easily accessible form on HepData. This project has received funding from the European Union's horizon 2020 research and innovation programme under grant agreement no 824093, aka STRONG 2020. CB acknowledges support from the Knut and Alice Wallenberg foundation, contract number 2017.0036.
\end{acknowledgement}

%%%%%%%% Bibliography 
\bibliographystyle{utphys}   
\bibliography{bibliography}

%%%%%%%%%%%%%%%%%%%%%%%%%%%%%%%%
% Appendices
%%%%%%%%%%%%%%%%%%%%%%%%%%%%%%%%
\newpage
\appendix
\section{Parameters}
\label{appendix:parameters}

\begin{table}[h!]
\centering
\begin{adjustbox}{max width=\textwidth}
\begin{tabular}{lcr}
        \hline
        \cline{1-3}
	Parameter name & Monash 2013 \cite{Skands:2014pea} & Used value  \cite{Bierlich:2015rha,Bierlich:2022ned}\\
        \hline
        \texttt{ColourReconnection:mode} & \texttt{-} & \texttt{1} \\
        \texttt{ColourReconnection:allowDoubleJunRem} & \texttt{-} & \texttt{off} \\
        \texttt{BeamRemnants:remnantMode} & \texttt{0} & \texttt{1} \\
        \texttt{Ropewalk:ropeHadronization} & \texttt{-} & \texttt{on} \\
        \texttt{Ropewalk:doFlavour} & \texttt{-} & \texttt{on} \\
        \texttt{Ropewalk:doShoving} & \texttt{-} & \texttt{on} \\
        \texttt{PartonVertex:setVertex} & \texttt{-} & \texttt{on} \\
        \texttt{StringFlav:probStoUD} & \texttt{0.217} & \texttt{0.2} \\
        \texttt{StringFlav:probQQtoQ} & \texttt{0.081} & \texttt{0.078} \\
        \texttt{StringZ:aLund} & \texttt{0.68} & \texttt{0.36} \\
        \texttt{StringZ:bLund} & \texttt{0.98} & \texttt{0.56} \\
        \texttt{StringFlav:mesonCvector} & \texttt{0.88} & \texttt{1.35} \\
        \texttt{StringFlav:probQQ1toQQ0join} & \texttt{0.5,0.7} & \texttt{0.0275,0.0275} \\
	& \texttt{0.9,1.0} & \texttt{0.0275,0.0275} \\
        \texttt{MultipartonInteractions:pT0Ref} & \texttt{2.28} & \texttt{2.15} \\
        \texttt{Ropewalk:beta} & \texttt{-} & \texttt{0.2} \\
        \texttt{Ropewalk:deltat} & \texttt{-} & \texttt{0.05} \\
        \texttt{Ropewalk:gAmplitude} & \texttt{-} & \texttt{0.0} \\
        \texttt{Ropewalk:tShove} & \texttt{-} & \texttt{0.1} \\
        \hline
        \cline{1-3}
\end{tabular}
\end{adjustbox}
%\end{center}
\caption{\label{table:pythia-parameters}Used parameters for \pythia extrapolation and estimation of kinematic charm mass. \pythia defaults are the so-called ``Monash tune'' \cite{Skands:2014pea}, and the used values are a combination of values from the two tunes in ref. \cite{Bierlich:2015rha}, carried out in ref. \cite{Bierlich:2022ned}, with the addition of a changed value of \texttt{StringFlav:mesonCvector}, which is not used elsewhere.}
\end{table}

\clearpage
\section{Additional plots and tables}
\label{appendix:plots}

\begin{figure}[!tbph]
\centering
\begin{minipage}[t]{1.00\textwidth}
\centering
\includegraphics[width=1.0\textwidth]{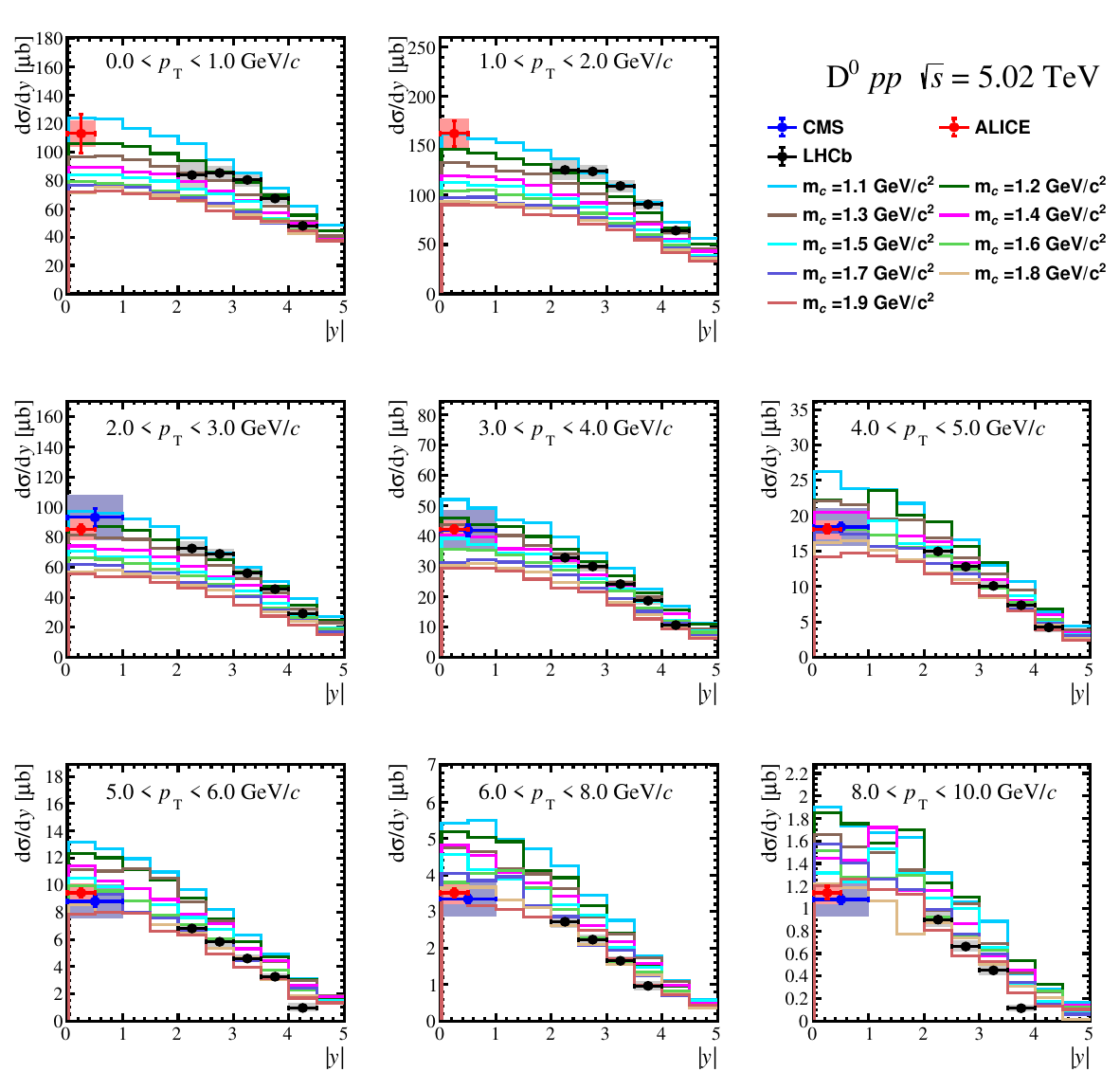}
\end{minipage}
	\caption{Measured $\Dzero$ cross section as a function of rapidity compared with \pythia simulations with a variation of the charm quark mass between $m_c=1.1$ and $1.9$ GeV/$c^2$, shown in different \pt intervals.
	}
\label{fig:pythia_mc_d0}
\end{figure}

\begin{figure}[!tbp]
\centering
\begin{minipage}[t]{1.00\textwidth}
\centering
\includegraphics[width=1.0\textwidth]{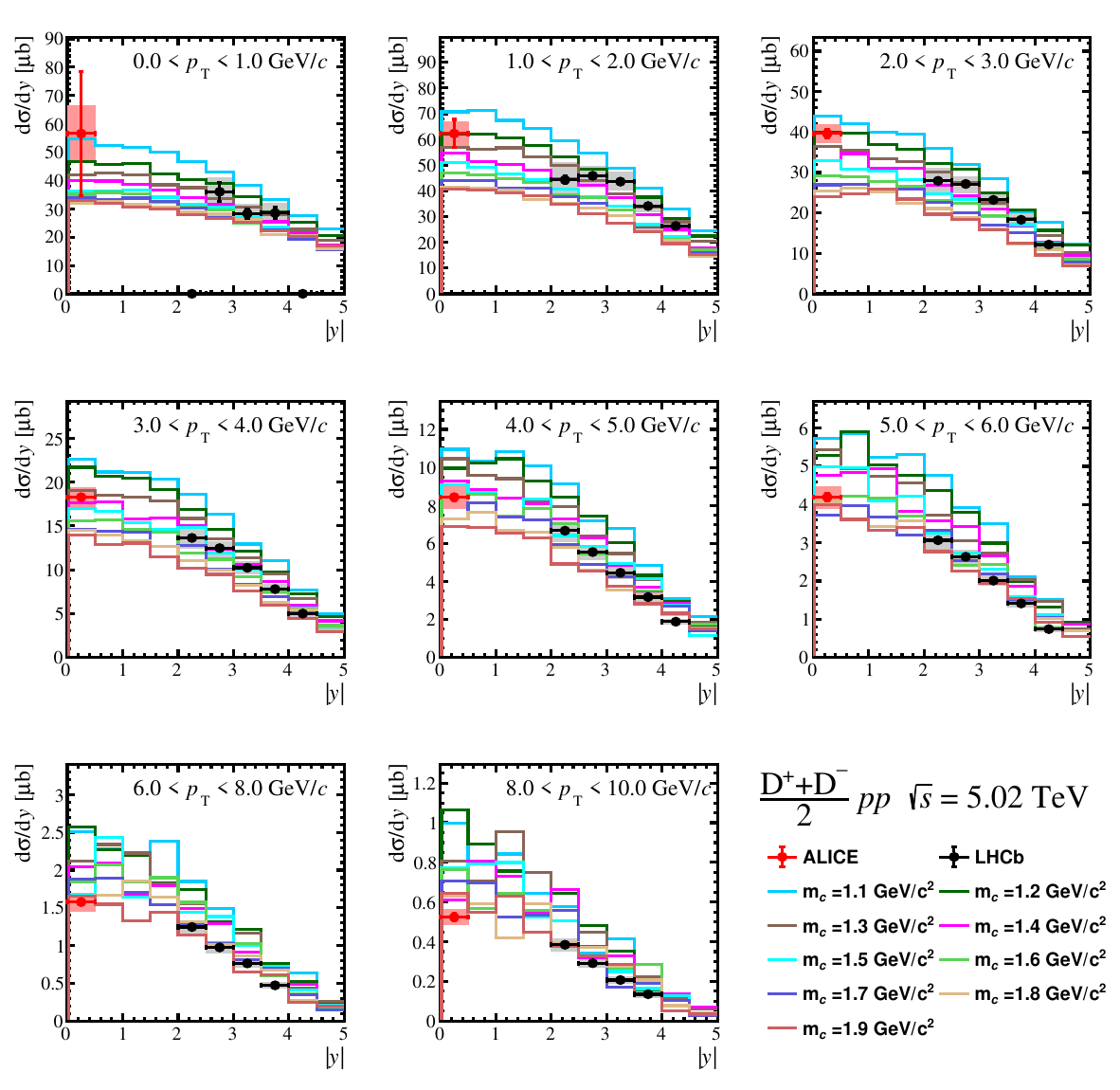}
\end{minipage}
	\caption{Measured $\Dplus$ cross section as a function of rapidity compared with \pythia simulations with a variation of the charm quark mass between $m_c=1.1$ and $1.9$ GeV/$c^2$, shown in different \pt intervals.
	}
\label{fig:pythia_mc_dp}
\end{figure}

\begin{figure}[!tbp]
\centering
\begin{minipage}[t]{1.00\textwidth}
\centering
\includegraphics[width=1.0\textwidth]{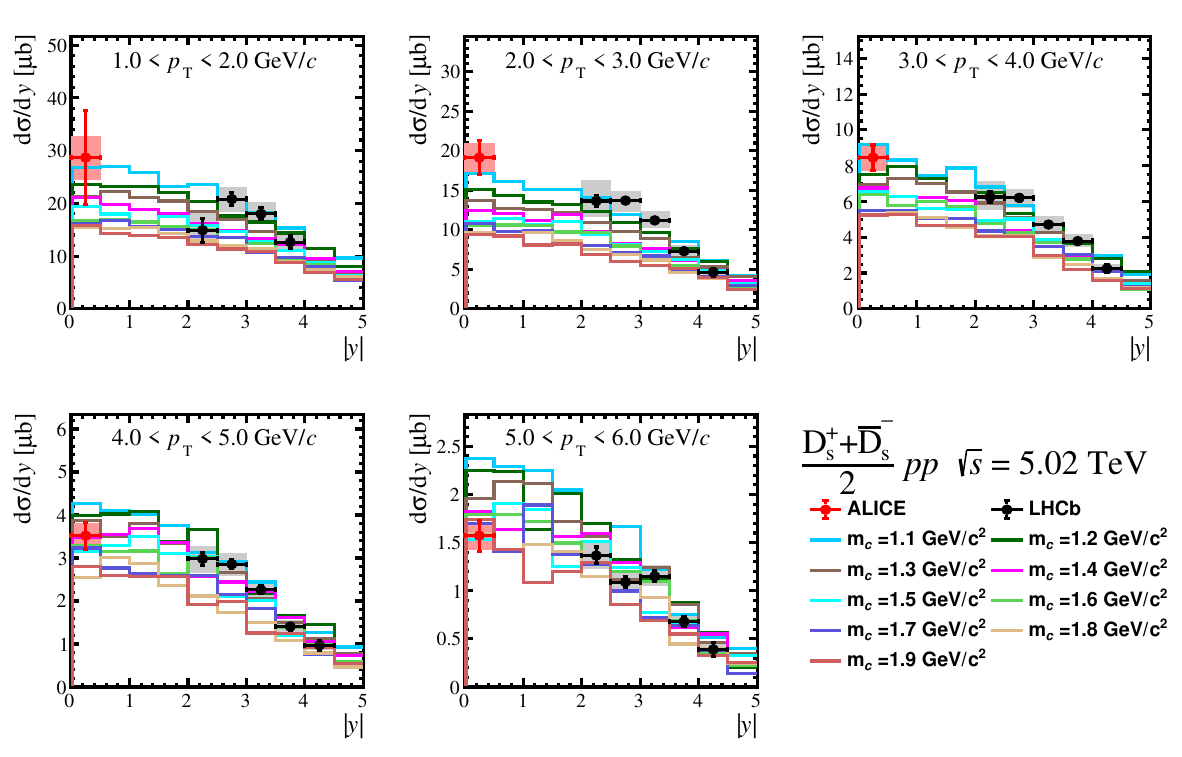}
\end{minipage}
	\caption{Measured $\Ds$ cross section as a function of rapidity compared with \pythia simulations with a variation of the charm quark mass between $m_{c}=1.1$ and $1.9$ GeV/$c^2$, shown in different \pt intervals.
	}
\label{fig:pythia_mc_ds}
\end{figure}

\begin{figure}[!tbp]
\centering
\begin{minipage}[t]{1.00\textwidth}
\centering
\includegraphics[width=1.0\textwidth]{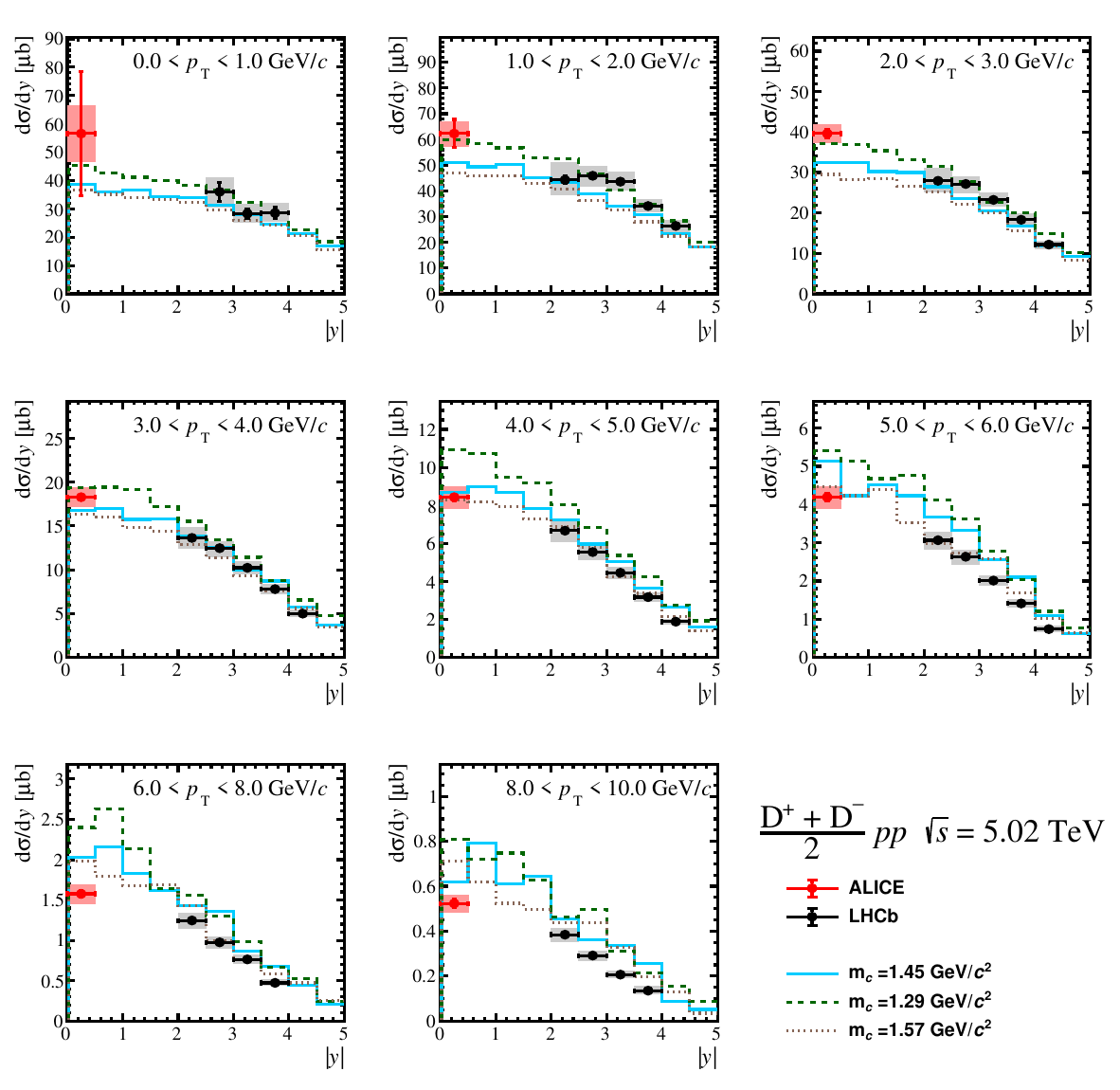}
\end{minipage}
	\caption{Measured \Dplus cross section as a function of rapidity compared with \pythia simulations with the bands corresponding to a 1$\sigma$ variation around the optimum.
	}
\label{fig:pythia_mc_dp_result}
\end{figure}

\begin{figure}[!tbp]
\centering
\begin{minipage}[t]{1.00\textwidth}
\centering
\includegraphics[width=1.0\textwidth]{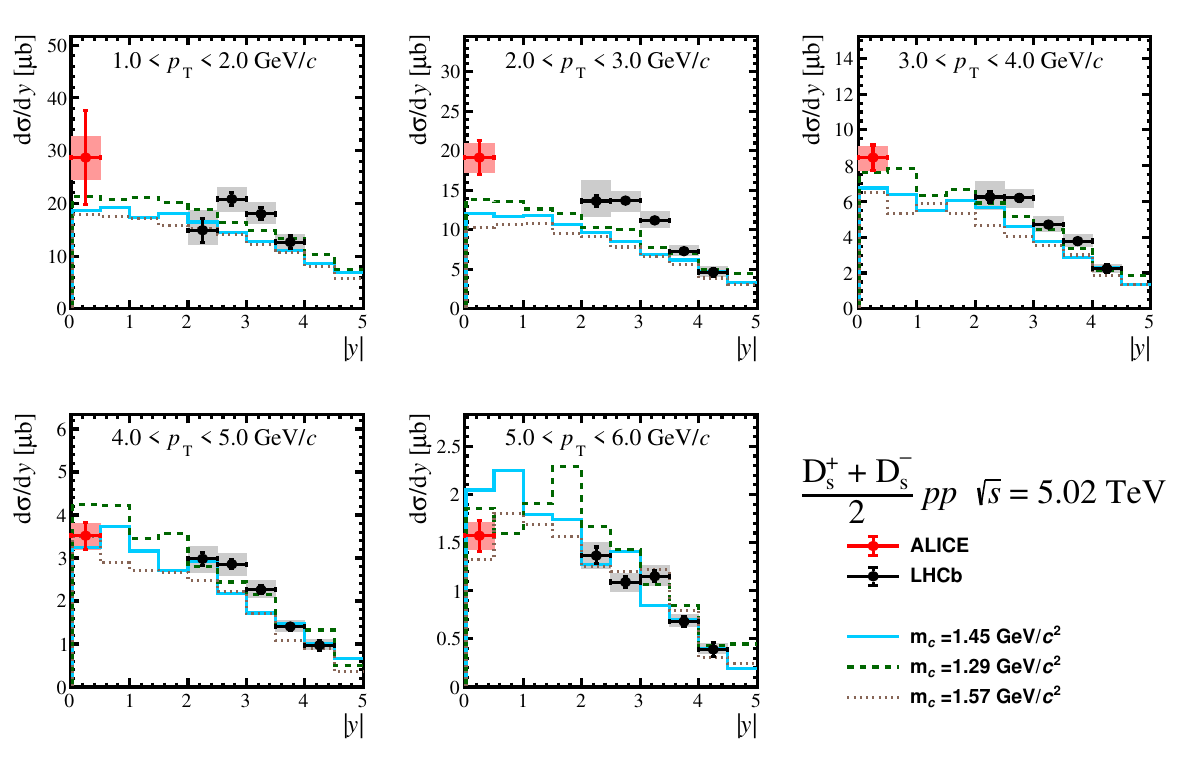}
\end{minipage}
	\caption{Measured \Ds cross section as a function of rapidity compared with \pythia simulations with the bands corresponding to a 1$\sigma$ variation around the optimum.
	}
\label{fig:pythia_mc_ds_result}
\end{figure}

\begin{figure}[!tbp]
\centering
\begin{minipage}[t]{1.00\textwidth}
\centering
\includegraphics[width=1.0\textwidth]{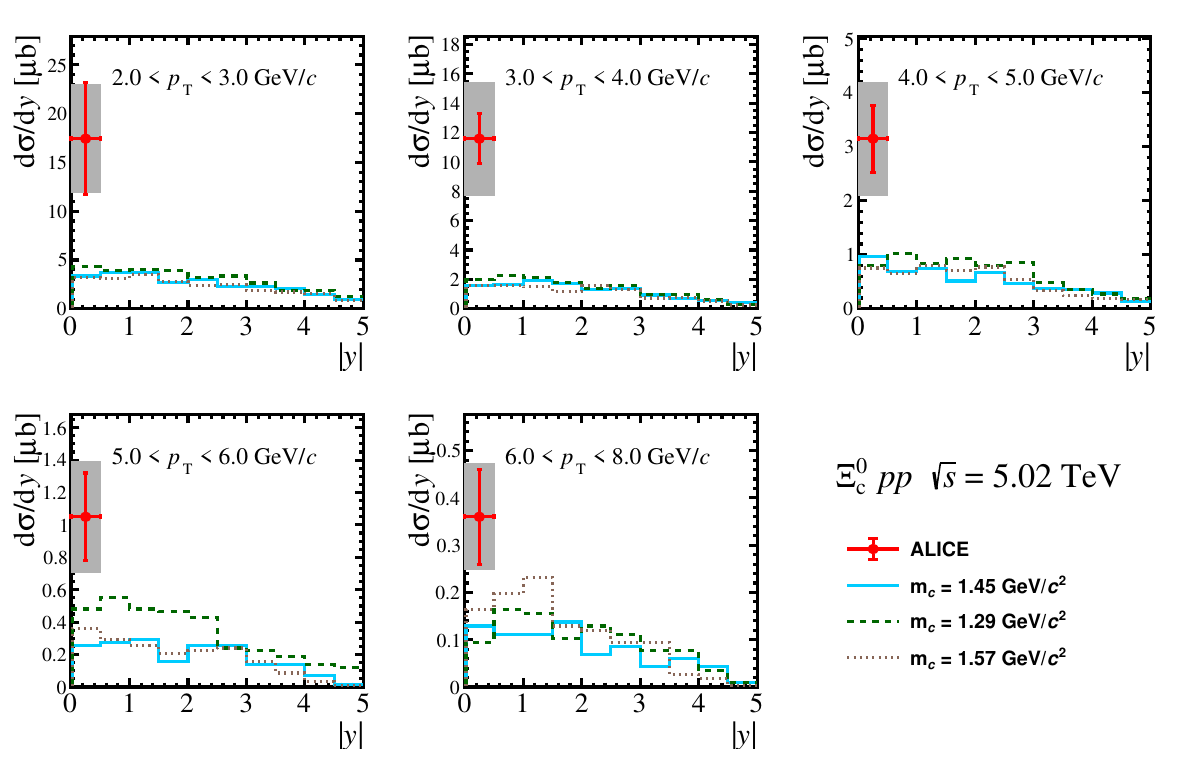}
\end{minipage}
	\caption{Measured \Xicz cross section as a function of rapidity compared with \pythia simulations with the bands corresponding to a 1$\sigma$ variation around the optimum.
	}
\label{fig:pythia_mc_xc_result}
\end{figure}

\newpage

\begin{table}[tpb]
    \centering
    \caption{Central values of the extrapolation factors in \pt used to complete the phase space within each experiment's rapidity coverage.}
    \begin{tabular}{ccc}
         &  ALICE, $|y| <  0.5$& LHCb, $2 < y < 4.5$\\\hline\hline
         \Dzero&  1& 1.007\\
         \Dplus&  1& 1.125\\
         \Ds&  1.335& 1.49\\
         \Lc&  1.001& --\\
         \Xicz&  2.707& --\\
    \end{tabular}
    \label{tab:ptextrapfacs}
\end{table}

\begin{table}[tpb]
    \centering
    \caption{Central values of the interpolation factors in rapidity used to convert the total hadron cross sections from the limited experimental rapidity coverage to $|y| < 8$.}
    \begin{tabular}{cc}
         & $|y| < 8$\\\hline\hline
         \Dzero& 1.853\\
         \Dplus& 1.856\\
         \Ds& 1.862\\
         \Lc& 7.959\\
         \Xicz& 7.402\\
    \end{tabular}
    \label{tab:rapinterpfacs}
\end{table}

\end{document}